\documentclass[runningheads]{llncs}
\usepackage{xcolor}
\usepackage[T1]{fontenc}
\usepackage{graphicx}
\usepackage{svg}
\begin{document}
\title{Multi-Agent Robotic Control with Onboard Vision-Language Models}
%
%
%
%
\author{ Kajetan Rachwał \inst{1,2}\orcidID{0000-0003-1524-7877} \and
Maciej Majek \inst{1}\orcidID{0009-0009-9541-8461} \and
Bartłomiej Boczek \inst{1}\orcidID{0009-0006-9097-9655} \and
Jakub Matejczyk \inst{1}\orcidID{0009-0007-4835-1829} \and
Dominik Matejkowski \inst{1}\orcidID{0009-0008-6803-8001} \and
Adam Dąbrowski \inst{1}\orcidID{0000-0002-2130-0577} \and
Tim Seyde \inst{3} \orcidID{0000-0001-9592-4465} \and
Alexander Amini \inst{3} \orcidID{0000-0002-9673-1267} \and
Maria Ganzha \inst{2}\orcidID{0000-0001-7714-4844}
}
\authorrunning{K. Rachwał et al.}
%
\institute{Robotec.AI, Warsaw, Poland \\ \email{\{name.surname\}@robotec.ai} \and Faculty of Mathematics and Information Science, Warsaw University of Technology, Warsaw, Poland \\ \email{maria.ganzha@pw.edu.pl} \and Liquid.AI, Cambridge, Massachusetts, USA \\ \email{\{name\}@liquid.ai}}
\maketitle              
\begin{abstract}
Vision Language Models (VLMs) and Vision Language Action (VLA) models have shown promise in robotic control. Yet, they face significant challenges regarding explainability, generalization, and compute requirements.
This paper presents a Multi-Agent System (MAS) architecture that addresses these limitations by deploying specialized agents on onboard hardware -- eliminating dependence on external compute.
The system controls a multi-purpose autonomous mobile manipulator in a simulated industrial warehouse, fulfilling five task categories: safety inspection, warehouse maintenance, warehouse search, package quality verification, and responding to human requests.
Compact VLMs (3-20B~parameters) are used throughout, with fine-tuning applied to improve package inspection accuracy.
A novel ``Megamind'' orchestration agent mitigates context retention issues inherent to long-horizon planning with smaller models.
The system was validated in a hardware-in-the-loop simulation using an AMD Ryzen~\texttrademark{} AI mini PC.
Results demonstrate that a fully onboard MAS architecture is a viable, cost-efficient alternative to cloud-dependent deployments, with strong potential for real-world transfer.
The simulation environment has been released as open source under the Apache~2.0 licence.

\end{abstract}
\section{Introduction}
Vision Language Models (VLMs) and Vision Language Action (VLA) models have been successfully utilized in the field of robotics in control tasks.
However, these solutions have faced certain drawbacks.
VLM based solutions typically require network connectivity and significant computing power in order to maintain an acceptable standard of operation.
VLAs sometimes face similar issues.
Their primary problem, however, is a general lack of explainability.
This raises safety and ethical concerns.
Some solutions to address these concerns exist, such as utilizing interpretability techniques on hidden layers or embodied chain-of-thought to intervene in steering~\cite{lee2025molmoact}. 
Other authors propose utilizing constraint functions~\cite{zhang2025}, or constraining deployment to ``safe'' robots~\cite{su2025}.
However, all of these methods suffer from problems with generalization capabilities on out-of-distribution data.

\begin{figure}[htbp]
  \centering
  \begin{minipage}[b]{0.49\textwidth}
    \includegraphics[width=\textwidth]{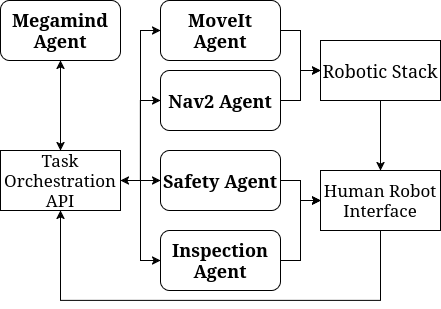}
    \caption{Deployed multi-agent architecture for vision-guided robotic inspection and safety monitoring.}
    \label{fig:archi}
  \end{minipage}
  \hfill
  \begin{minipage}[b]{0.4\textwidth}
    \includegraphics[width=\textwidth]{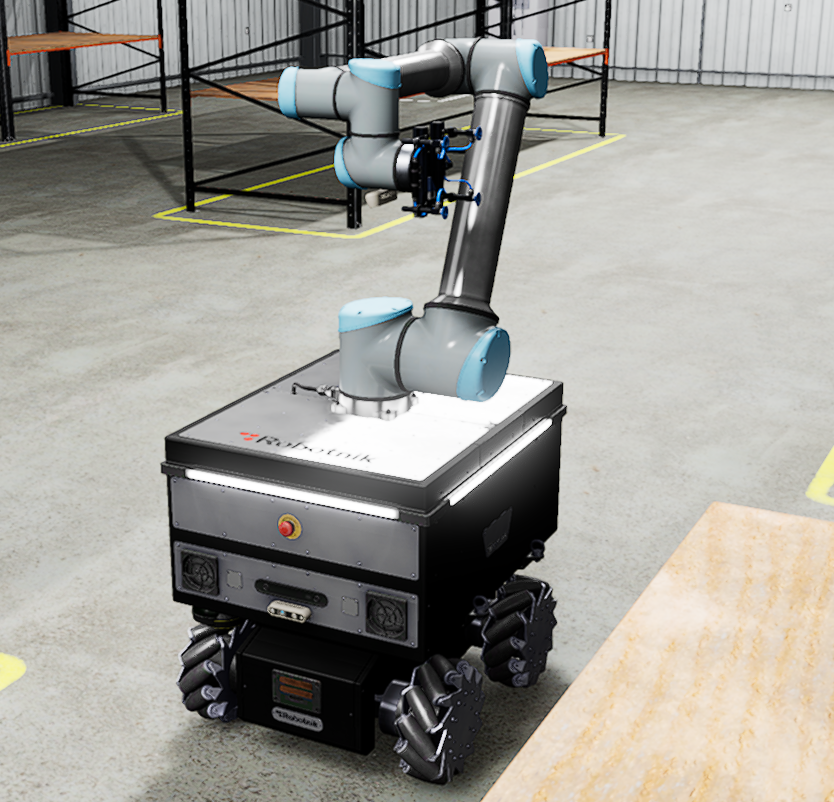}
    \caption{Robot platform in the simulated environment.}
    \label{fig:env}
  \end{minipage}
\end{figure}

The goal of this article is to demonstrate how these problems could be addressed utilizing Multi-Agent-System-based architecture (MAS).
To demonstrate this, a multipurpose autonomous mobile manipulator has been deployed.
It's controlled by a MAS~(Fig.~\ref{fig:archi}) running fully on the robot's hardware.
The system is capable of real-time response and operation in dynamic industrial environments.
It has been tested in a hardware-in-the-loop simulation of a warehouse~(Fig.~\ref{fig:env}).
The simulated robot platform uses RB-KAIROS+ coupled with an AMD Ryzen~\texttrademark{} AI mini PC for inference of the VLMs employed by the system's agents.
The VLM deployed on the robot's hardware was LFM2-VL-3B~\cite{amini2025lfm2}.
Utilization of hardware-in-the-loop allows for feasibility assessment of using small locally hosted VLMs for robot control.


\subsection{Objectives}\label{sec:objectives}

The demonstrated setup was developed to fulfil the following tasks: Safety inspection~(\textbf{T1}), warehouse maintenance~(\textbf{T2}), searching for objects~(\textbf{T3}), package quality verification~(\textbf{T4}), and responding to human requests~(\textbf{T5}).

Tasks \textbf{T1} and \textbf{T2} are performed continuously in the background, while tasks \textbf{T3-T5} are performed periodically or upon request.
This diverse set of objectives would require a varied and extensive context for a single VLM to perform.
When using a smaller model, this requirement can become detrimental as crucial information may slide out of the context window.
A possible solution to this problem is the use of memory storage techniques.
However, memory-based solutions may still cause context scaling issues as the variety of tasks grows.
In this work, we have opted for a MAS, with specialized agents focusing on different tasks.

The following section discusses the problems that the proposed architecture solves.
In section~\ref{sec:architecture}, the proposed architecture is described in detail.
During the demonstration (section~\ref{sec:demonstration}) the participants will be able to interact with the simulated robot by giving it commands.
The utility and quality of the solution, as well as further research and development directions, are then concluded in section~\ref{sec:conclusions}.

\section{Main Purpose}

The demonstration intends to showcase the capabilities of a comprehensive MAS fulfilling a broad set of tasks.
It presents the capabilities and limitations of such systems running entirely on onboard compute platforms.
To make this possible, compact LLMs (3-20B parameters) have been used.
This setting is highly relevant to the application of flexible robotics in small- to medium-sized enterprises.
Deployments that require the use of external compute platforms (either edge- or cloud-based) can incur higher operational or initial costs -- both in the form of electricity and hardware.

This contribution also intends to provide an advanced platform for testing and development of robotic agentic systems.
The simulation, including a configurable warehouse, packages, and the robot platform, has been made fully open source under the Apache~2.0 license.
It is fully integrated with ROS~2 and RAI~\cite{rachwal2025}, providing ready-to-use software that enables research into agentic systems for mobile manipulators.



\section{Demonstration}

\subsection{Architecture}\label{sec:architecture}

We designed a multi-agent system (MAS) to address the objectives in Section~\ref{sec:objectives} and implemented it using the RAI framework~\cite{rachwal2025}. Figure~\ref{fig:archi} shows the resulting architecture. The system combines a supervisory planning agent, vision-based inspection and safety agents, and low-level control agents for navigation and manipulation. Users interact with the MAS through a Human-Robot Interface (HRI), which provides a graphical interface with a live video feed and status updates.

\paragraph{\textbf{Vision Language Model.}}
The Megamind, Inspection, and Safety agents use a shared 3B-parameter VLM (LFM2-VL-3B~\cite{amini2025lfm2}) for tasks \textbf{T1} and \textbf{T4}.
During development, we found that the base checkpoint struggled with package-quality inspection in task \textbf{T4}.
To address this limitation, we fine-tuned the model on a simulation-derived dataset.
Each image was annotated with three attributes: \texttt{anomaly} (True/False), \texttt{box\_condition} (good/damaged), and \texttt{safety\_hazard} (True/False).
We then conditioned Qwen3-256B-A22B on these annotations, as a teacher model, to generate synthetic training data of reference natural-language descriptions.
The VLM was fine-tuned for one epoch on the resulting dataset, producing a single checkpoint shared across all visual tasks in the demonstration.
The fine-tuning increased box condition annotation accuracy from 76.7\% to 91.5\% (F1-score: 0.755 to 0.915).
The most significant improvements were the increase in true negatives (from 54\% to 95.5\%) and the decrease in false positives (from 46\% to 4.5\%).
The final checkpoint is released under the LFM Open License\footnote{\url{https://huggingface.co/LiquidAI/LFM2-VL-3B/blob/main/LICENSE}}.

\paragraph{\textbf{Megamind Agent.}}
Compact LLMs can struggle with context retention during long-horizon planning and execution.
To mitigate this limitation, we introduce \emph{Megamind}, a supervisory agent organized as a two-state self-feedback loop.
In the planning state, Megamind selects the next task based on the current robot state and task queue, then delegates it to the appropriate sub-agent.
Once the sub-agent returns its output, Megamind transitions to the analysis state, where it determines whether the task was completed successfully.
If recovery is required, Megamind generates corrective steps and appends them to the task queue before returning to the planning state.
This explicit control loop externalizes task progress, delegation, and recovery, enabling compact LLMs to support multi-step robotic tasks.

\paragraph{\textbf{Inspection Agent.}}
The Inspection agent monitors the front-facing camera feed for warehouse anomalies, including misplaced boxes, floor obstructions, and spills.
It follows a two-stage workflow.
First, the VLM audits each image, generates a concise natural-language description, and flags up to three potential issues.
Second, the flagged issues are categorized as \texttt{box} for graspable boxes in the aisle, \texttt{trash} for graspable litter, \texttt{other} for non-manipulable obstacles that should only be reported, or \texttt{nothing} when no relevant issue is present.
For graspable objects, the agent obtains the 3D pose from simulation ground truth.
It outputs a structured JSON object containing the issue type, pose, and short description, which is passed to Megamind for action selection.

\paragraph{\textbf{Safety Agent.}}
The Safety agent performs continuous regulatory-compliance monitoring using the front-facing camera feed.
Detected violations are logged and reported through the HRI for operator review.
Compliance assessment is implemented as a three-stage pipeline: vision, retrieval, and reasoning.
In the \emph{vision} stage, the fine-tuned VLM describes the scene and identifies up to three potential safety hazards, returning a structured observation with an explicit list of anomalies.
The model is instructed to report only hazards supported by clear visual evidence. 
In the \emph{retrieval} stage, the agent queries a vector database for each identified hazard.
The database is constructed offline from OSHA 29~CFR~1910 regulations\footnote{\url{https://www.osha.gov/laws-regs/regulations/standardnumber/1910}}, filtered to warehouse-relevant sections, and indexed using Qwen3-Embedding-0.6B~\cite{yang2025qwen3}.
Retrieved candidates are re-ranked using Qwen3-Reranker-0.6B, and only documents above a relevance threshold are retained, with at most three excerpts per hazard.
In the \emph{reasoning} stage, the VLM receives the original image, the scene description, and the retrieved regulatory excerpts for each hazard.
It determines whether the hazard constitutes a violation and, if so, outputs the applicable regulation identifiers, a supporting excerpt, a severity level (\texttt{LOW}, \texttt{MEDIUM}, \texttt{HIGH}), and a brief rationale.
Processing hazards independently keeps the context window small for the quantized checkpoint running on edge hardware.
To reduce inference, consecutive frames with high visual similarity, as measured by Structural Similarity Index Measure, are skipped.

\paragraph{\textbf{MoveIt and Nav2 Agents.}}
The MoveIt and Nav2 agents provide low-level control for the manipulator and mobile platform, respectively.
We implemented corresponding RAI tools that interface with MoveIt~2\footnote{\url{https://moveit.ai/}} and Nav2\footnote{\url{https://docs.nav2.org/}}, two ROS~2-based stacks for manipulation and navigation.
Both agents execute commands issued by Megamind and report status updates through the HRI.

\subsection{Demonstration}\label{sec:demonstration}

The demonstration will run across three nodes: an O3DE simulation host, a Hardware-in-the-Loop (HIL) compute node running the full agent stack and local inference on an AMD Ryzen~\texttrademark{} AI mini PC (128~GB unified RAM, 96~GB allocated to the GPU), and a the HRI display.
Attendees will be able to interact with the running system by issuing natural language commands through the Qt-based HRI.\penalty-10000

The demonstration will cover four task categories.
First, operator will direct the robot to move packages between named racks and tables~(\textbf{T5}).
Second, the inspection agent will inspect a floor anomaly using a two-stage VLM pipeline (a free-text descriptive query followed by a structured-output classification call) and report the object type and pose to the orchestrator~(\textbf{T3}).
Anomaly tasks enter the high-priority queue, so the orchestrator will preempt the active user task and dispatch the robot to relocate the object to an inspection area or dispose of it in a bin.
Third, the housekeeping executor will correct misaligned packages on a designated rack~(\textbf{T2}).
Fourth, the safety agent will cross-reference the camera frame against a FAISS vector database of OSHA regulations and report the applicable clause to the operator~(\textbf{T1}, \textbf{T4}).

\section{Conclusion}\label{sec:conclusions}

This work shows that a Multi-Agent System using compact, locally hosted Vision-Language Models can effectively control a versatile autonomous mobile manipulator across diverse industrial tasks. Hardware-in-the-loop simulation validated feasibility before physical deployment.

Robotic tasks pose unique challenges -- real-time reactivity, heterogeneous sensors, and certifiable, interpretable control -- that reinforcement-learned models and generalist policies struggle to meet.
The MAS approach addresses these via mature stacks (ROS~2, MoveIt~2, Nav~2), modular retooling, and heterogeneous compute across agents.

Fine-tuned compact models were cost-efficient and quick to deploy; knowledge distillation from a large teacher model delivered strong task-specific performance at far lower inference cost.
Integrating established tools (FAISS, Qwen3-Embedding, Qwen3-Reranker) reduced development overhead while maintaining high capability.

Success across all five task categories justifies real-robot trials and opens research avenues for agentic systems in flexible robotics for small- and medium-sized enterprises.


\begin{credits}
\subsubsection{\ackname} The authors want to thank AMD for providing funding and hardware used in the development of this demonstration. 
\subsubsection{\discintname}
The authors have no competing interests to declare that are relevant to the content of this article.
\end{credits}
%
%
%
\bibliographystyle{splncs04}
\bibliography{refs}
\end{document}